\documentstyle[twoside,fleqn,espcrc2]{article}

\newcommand{\beq}{\begin{eqnarray}}
\newcommand{\eeq}{\end{eqnarray}}
\newcommand{\nn}{\nonumber\\}

\newcommand{\AmS}{{\protect\the\textfont2
  A\kern-.1667em\lower.5ex\hbox{M}\kern-.125emS}}
\newcommand{\prd}{Phys.~Rev.~D\ }
\title{Pre-Big Bang Cosmology and Quantum Fluctuations 
}
\author{A. Ghosh,\address{104 Davey Lab, Penn State,  
        University Park, PA 16802, USA}
         G. Pollifrone\address{Astronomy Unit, School of 
Mathematical Sciences,\\Queen Mary \& 
Westfield College, Mile End Road, E1 4NS London,
      United Kingdom} and 
        G. Veneziano\address{Theory Division, CERN, 
CH-1211 Geneva 23, Switzerland}}
\begin{document}
\begin{abstract}
The quantum fluctuations of  a homogeneous, isotropic, open 
pre-big bang model are discussed.
By solving exactly the equations for tensor and scalar perturbations 
we find that particle production is negligible during the 
perturbative Pre-Big Bang phase.
\end{abstract}
\maketitle
\section{INTRODUCTION}
In the framework of string theory, the Pre-Big Bang scenario 
\cite{PBB,veneziaerice} 
provides an alternative to the standard inflationary paradigm. 
In the Pre-Big Bang model \footnote{
An updated collection of papers on the PBB scenario is available at 
{\tt http://www.to.infn.it/${\sim}\!\!\!$ gasperin/}.} 
the inflationary solutions, driven by 
the kinetic energy of the dilaton, emerge naturally via the 
duality symmetries of string theory. 

It has, however,  been argued \cite{KLB} that, 
even though the two classical moduli of the 
open (${\cal K}=-1$), homogeneous and isotropic solution 
\cite{CLW,TW} lie  deeply inside the perturbative regime,  the
vacuum quantum fluctuations drastically modify 
the classical behaviour  
preventing the occurrence of an appreciable amount of inflation.

Quantum fluctuations in a non-spatially 
flat Universe are considerably harder to 
study than in the flat case \cite{MFB,GMST}. 
In Ref. \cite{GPV} we thoroughly studied the  
quantum fluctuations around the  ${\cal K}=-1$ solution \cite{CLW,TW}.
In that work we showed that the perturbation 
equations can be {\it exactly} integrated in terms of standard 
hypergeometric functions. We  found that particle production 
(i.e. the amplification of vacuum fluctuations) is strongly 
suppressed at very early times and remains small through the whole
perturbative PBB phase, and hence, does not impede the occurrence 
of PBB inflation.
\section{THE SECOND-ORDER ACTION}
The (string-frame) open  homogeneous, isotropic 
PBB-type solution was first found in \cite{CLW} and
then rederived and discussed in \cite{TW}. The solution contains
two arbitrary moduli, $L$ and $\phi_{\rm{in}}$, reflecting the
symmetries of the classical equations under a constant shift of 
the dilaton and a constant rescaling of the metric.
These two parameters are to be chosen appropriately (see Refs.  
\cite{KLB,TW,inhomog,maggiore,BMUV,MOV,BDV,gasper}) in order to ensure
the occurrence of a sufficient amount of PBB inflation. Such 
a solution describes a universe which is almost trivial (Milne-like) 
at $ \eta\to-\infty$ and inflating in $-\infty<\eta<{\cal O}(1)$, 
having an initial curvature ${\cal O}(L^{-2})$ and coupling 
${\cal O}(\exp(\phi_{\rm
{in}}/2))$, until it enters the strong curvature and/or strong coupling
regime at $\eta \sim \eta_1$. The critical value $\eta_1$ is easily
determined in terms of the integration constants $L$ and $\phi_{\rm
{in}}$ as $(- \eta_1) = 
\max\,( e^{\phi_{\rm{in}}/\sqrt{3}} , (\ell_s/L)^{1 + 1/\sqrt{3}} )$.

It is well known \cite{PBB} that studying perturbations is 
technically simpler in the so-called Einstein-frame which is 
related to the string-frame by a conformal transformation. The
action is
\beq
{S^{(E)}}={1\over 2\ell_P^2}\int d^4x\sqrt{-g}\left(R(g) 
-{1\over 2}(\partial\phi)^2
\right)\;,
\label{Eaction}
\eeq
where $\phi_{\rm{today}}$ is the present value of the dilaton, 
$\ell_P \equiv \sqrt{8 \pi G } = \exp(\phi_{\rm{today}}/2)  
\ell_s \sim 0.1 \ell_s$ refers to the present value of the 
Planck-length with $ \hbar =1$. Usually 
one computes perturbations in the Einstein frame and 
then transforms the results  back  to the string 
frame for a physical interpretation.
 
The ${\cal K}=-1$ solution is:
\beq
a(\eta)&=&\ell~(-\sinh\eta\cosh\eta)^{1\over 2}\nn
\phi(\eta)&=&-\sqrt 3~\ln(-\tanh\eta)+\phi_{\rm{in}} \; , \; \;  
\eta < 0 \; ,
\label{esol}
\eeq
where the modulus $\ell$ is given by 
$\ell^2=L^2 \exp (\phi_{\rm{today}} - \phi_{\rm{in}})$.
Generic perturbations are defined by
\beq
g_{\mu\nu}=g^{(0)}_{\mu\nu}+\delta g_{\mu\nu}\;\; , \;\;
\phi=\phi^{(0)}+\delta\phi 
\label{genpert}
\eeq
where superscript $(0)$ refers to the background solution 
and we shall use isotropic-spatial coordinates.
\section{QUANTUM FLUCTUATIONS}
\subsection{Tensor perturbations}
Since the tensor metric perturbations are  automatically  
gauge-invariant and decoupled from the scalar perturbations, they are
easier to study. They are defined as
\beq
\delta g_{\mu\nu}^{\rm {(T)}}={\rm{diag}}(0,a^2 h_{ij})\;,
\label{tensor}
\eeq
where the symmetric three-tensor
$h_{ij}$ satisfies the transverse-traceless (TT) conditions.

 We then  find:
\beq
\delta^{(2)} S^{(T)} &=& 
{1\over 4 \ell_P^2}\int d^4x\sqrt{\gamma}\; a^2\left(
{h'^{ij}}h'_{ij} \right.\nn
&-&\left. \nabla^l h^{ij}\nabla_l h_{ij}-2{\cal K}h^{ij}h_{ij}\right)
\; .
\label{tensoraction}
\eeq
By expanding the tensor perturbations 
in TT tensor-pseudospherical harmonics (as ${\cal K}=-1$) \cite{LKGS},   
we eventually get the simple equation
\beq
{u}_{nlm}''+\left(n^2+{1\over 12}\phi'^2\right){u}_{nlm}=0\; ,
\label{hmotion}
\eeq
where ${u}_{nlm}\equiv a {h}_{nlm}$ is the canonical variable
of perturbation. For the background (\ref{esol})  Eq. (\ref{hmotion}) 
can be exactly solved in terms of the standard hypergeometric function  
\cite{ABST} as
\beq
&&{u}_N(\eta)=
C_1  \;[{\rm{csch}}^2(2\eta)]^{-{in\over 4}}\times\nn
&&F
\left[{1-in\over 4},{1-in\over 4},{2-in\over 2},-
{\rm{csch}}^2(2\eta)\right]
\nn &&+C_2\;{\rm c.c.}\; ,
\label{hypergeomten}
\eeq
where $N$ stands for the collection of indices $(nlm)$ and $C_{1,2}$ are
(classically arbitrary) integration constants.
At early times,
$n^2 \gg \phi'^2$, and thus $u$ is a free canonical field. Hence, imposing 
the standard commutation relations, as $\eta \rightarrow - \infty$, we get 
\beq
{u}_{N}(\eta) \rightarrow
 {u}_{N}^{-\infty}(\eta) \equiv {2\ell_P \over \sqrt{n}}e^{-in\eta}.
\label{hearly}
\eeq
Since 
$F[a,b,c,0]=1$, Eq. (\ref{hearly})  fixes the integration constants  
as $|C_1|=2\ell_P/\sqrt{n}, C_2 =0$.
The deviation from a trivial  plane-wave behaviour can easily be computed
 from the small argument limit of $F$. We find
\beq
u_N(\eta) = u_N^{-\infty}(\eta)\left(1+\alpha_n\,
e^{4\eta-i\beta_n}\right)\; ,
\label{earlyh}
\eeq
where  $\alpha_n,\beta_n$
are $n$-dependent constants fixed from the Taylor expansion of the
hypergeometric function. It is worth noting that the correction
to the vacuum amplitude dies off as
$e^{4\eta}$, i.e. as $t^{-4}$ in terms of cosmic time $t \sim -e^{-\eta}$.

We can also estimate the behaviour of the solution near the
singularity, i.e. for $\eta \rightarrow 0$.
By virtue of the small $\eta$ behaviour
$a \simeq \ell|\eta|^{1/2}$,  we find
\beq
|h_N|\simeq 2 \sqrt{{2 \over \pi }} {\ell_P \over \ell}
\sqrt{\coth{\left(n\pi\over 2
\right)}}\ln|\eta|\;.
\eeq
We shall come back to this result after deriving a similar
 expression for scalar perturbations.

\subsection{Scalar perturbations}
Consider now scalar metric-dilaton perturbations (\ref{genpert}).  
The scalar part of metric perturbations 
is defined by \cite{MFB}
\beq
\delta g_{\mu\nu}^{\rm {(S)}}
\equiv -a^2(\eta)\left(\matrix{2\varphi&\nabla_iB
\cr  \nabla_iB&2
(\psi\gamma_{ij}+\nabla_i\nabla_j E)\cr}\right) .
\label{scalarpert}
\eeq

In the  second-order action the variables $B,\varphi$ 
are Lagrange multipliers, providing two constraints. 
We can introduce the gauge-invariant
variable $\Psi$  by (see Ref. \cite{GMST})
\beq
\Psi={4\over\phi'}\big[\psi+{\cal H}(B-E')\big] \; ,
\label{gaugepsi}
\eeq
 and, after using the constraints,  the action  
 reads
\beq
&&\delta^{(2)}S^{(S)}={1\over 2\ell_P^2}\int d^4x\;a^2\sqrt \gamma\times\nn
&&(\nabla^2+3{\cal K})\Psi\left[
\partial^2_\eta-\nabla^2+2({\cal H}'+{\cal K})\right]\Psi .
\label{scalaraction}
\eeq
 One can now make use of the constraints to eliminate the variable
$(B-E')$ from the action (\ref{scalaraction})
in terms of $\varphi,\psi$ and
$\delta\phi$.
The latter variables are
 not independent either, being related
by a linear combination of the two constraints.
After its implementation the action (\ref{scalaraction}) contains only
true degrees of freedom.

As was in the case of  tensor perturbations,
we introduce a canonical field $\Psi_c$ and expand it  as
\beq
\Psi_c \equiv a \Psi =
\int\!dn\sum_{l=0}^{\infty}\sum_{m=-l}^{l}\Psi_{nlm}
(\eta)Q_{nlm}({{\bf x}}),
\eeq
where $Q_{nlm}({{\bf x}})$ are the scalar pseudospherical harmonics
\cite{LKGS}.
We then get a simple equation for 
$\bar\Psi_N \equiv \sqrt{n^2+4}\Psi_N$, namely 
\beq
\bar\Psi_N''+ (n^2- {1\over 4} \phi'^2)\bar\Psi_N=0\; ,
\label{psibarmotion}
\eeq
where we must
impose, as $\eta \rightarrow - \infty$,
\beq
\bar\Psi_N(\eta) \rightarrow \bar\Psi_N^{-\infty}(\eta)\equiv{\ell_P\over \sqrt{n}}e^{-in\eta}\; .
\label{normapsi}
\eeq
Eq. (\ref{psibarmotion})  can again be transformed (for the background
(\ref{esol})) into a hypergeometric equation. We hence find 
 \beq
&&\bar\Psi_N(\eta)=\tilde{C_1}\;[{\rm{csch}}^2(2\eta)]^{-{in\over 4}}\times\nn
&&F\,[{-1-in\over 4},{3-in\over 4},{2-in\over 2},-{\rm{csch}}^2(2\eta)]
\nn &&+\tilde{C_2}\; {\rm c.c.}\; ,
\label{hypergeompsi}
\eeq
where, as before, we have to take
$|\tilde{C_1}|=\ell_P/\sqrt{n}, \tilde{C_2}=0$. Corrections to the free
 plane wave can be easily computed
and, again, are suppressed by four powers of $1/t$:
\beq
\bar\Psi_N(\eta)=\bar\Psi_N^{-\infty}(\eta)\left(1+\tilde\alpha_n\,
e^{4\eta-i\tilde\beta_n}\right)\; ,
\label{farpast}
\eeq
where $\bar\Psi_N^{-\infty}$ is given by (\ref{normapsi}) 
and
$\tilde\alpha_n,\tilde\beta_n$
are $n$-dependent constants fixed from the expansion of the
hypergeometric function.

Estimating  the behaviour of (\ref{hypergeompsi}) near $\eta\simeq
0$  \cite{ABST},
we obtain:
\beq
&&|\bar\Psi_N|\simeq \ell_P\sqrt{n^2+1\over 2\pi}\sqrt{\coth{\left(n\pi\over 2
\right)}}\times\nn
&&\left(-|\eta|^{3/2}\ln|\eta|+{2\over n^2+1}|\eta|^{-1/2}\right)\; .
\label{psinear0}
\eeq
\section{CONCLUSIONS}
Let us choose the off-diagonal gauge \cite{Hwang,BGGMV},  defined by  
setting  $\psi=E=0$   in (\ref{scalarpert}). 
By using Eq. (\ref{gaugepsi}) one  can  
reconstruct  the scalar field fluctuation $\delta\phi$ from
$\Psi$ as 
\beq   
\delta\phi=\Psi'+{{\cal K}-{\cal H}'\over
{\cal H}}\Psi\; ,
\label{relphipsi}
\eeq
implying that $\delta\phi$ represents, in this gauge, a gauge-invariant
object.

In the
presence of spatial curvature, the field  
 $v=a\delta\phi$  plays 
the role of the canonical
field in the far past, when $\eta$ is large and negative. 
Eq.  (\ref{relphipsi}) tells us that the behaviour  of $v$ in the far past
follows directly from that of $\bar\Psi_N$, given in Eqs.
(\ref{normapsi}) and (\ref{farpast}):
\beq
v^{-\infty}(\eta)\equiv{\ell_P\over \sqrt{n}}\sqrt{2-in\over 2+in}
e^{-in\eta}\; .
\label{vnormalization}
\eeq
Corrections to (\ref{vnormalization}) are again suppressed as $t^{-4}$
\beq
v(\eta)=v^{-\infty}(\eta) \left(1+\hat\alpha_n\,
e^{4\eta-i\hat\beta_n}\right)\; ,
\label{earlyv}
\eeq
where ${\hat\alpha_n},\hat\beta_n$ are $n$-dependent constants.
 
The behaviour of $\delta\phi$  near $\eta \simeq 0$ is :
\beq
|\delta\phi_N|\simeq{\ell_P\over\ell}\sqrt{n^2+1\over 2\pi}
\sqrt{\coth({n\pi\over 2})\over n^2 +4}\ln|\eta|\; .
\eeq
Lastly, let us compare the energy contained in the
quantum fluctuations of the dilaton and that in the classical solution 
near the singularity. Note that the expansion (\ref{psinear0})
can be trusted
only up to some maximum $n$ for which $1\ll n_{\rm{max}}\sim 1/|\eta|$.
Consequently, the ratio of the kinetic energy densities
near $|\eta|\simeq 0$ (up to constant prefactors of ${\cal O}(1)$) becomes
\beq
{{\cal E}_{\rm Q}\over{\cal E}_{\rm C}}=
{\int d^3x\sqrt{\gamma}\;
a^2(\delta\phi')^2
\over
\int d^3x\sqrt{\gamma}\;
a^2{\phi'}^2}
\simeq
{\ell_P^2 \over \ell^2} \int^{n_{\rm{max}}}{dn\over n}\; n^3 \; .
\label{ratio}
\eeq
We can express the above result in terms of the value of
 the physical Hubble parameter
 $H(\eta)\equiv {\cal H}/a$ at horizon crossing
of the scale $n$, $H_{\rm {HC}}(n)$, i.e.  
\beq
H_{\rm {HC}}(n) \sim {1 \over \eta a}(\eta \sim 1/n) 
\sim n^{3/2}/\ell \;.
\label{HHC}
\eeq
Thus Eq. (\ref{ratio}) takes the suggestive form
\beq
{{\cal E}_{\rm Q}\over{\cal E}_{\rm C}}= 
\ell_P^2  \int^{n_{\rm{max}}}{dn\over n}\; H^2_{\rm {HC}}(n) \; .
\label{horx}
\eeq
In order to draw  physical conclusion  we should transform 
the results back 
to the string frame. 
However, in our case, this is hardly
necessary. As far as  the importance of vacuum
 fluctuations is concerned, as $\eta \rightarrow 0$,
the final result (\ref{horx}) expresses the relative
importance of quantum and classical fluctuations near the singularity in
terms of a frame-independent quantity: the ratio of the effective
Planck length to the size of the horizon. Since, by definition of the
perturbative dilaton phase, the Hubble radius is always larger 
than the string scale,
the relative importance of quantum fluctuations is always
bounded by the ratio $\ell_P / \ell_s$ which is always
less than one in the perturbative phase.

 Let us now come to the more subtle issue of
the far-past behaviour of tensor and
scalar quantum fluctuations. 
Computations may be done in either frame,
since the dilaton is approximately constant in the far past. Our results,
expressed in Eqs. (\ref{earlyh}) and (\ref{earlyv}), show
 that corrections to the trivial quantum fluctuations are of relative order
 $e^{4\eta} \sim t^{-4}$, i.e. of order $t^{-3}$ relative to the 
(homogeneous) classical 
 perturbation. This suggests that quantum effects do not
modify appreciably classical behaviour in the far past, 
contrary to the claim of  \cite{KLB}. 
This result is also supported 
 by the structure of the superstring one-loop effective-action 
 (which is well-defined thanks to the string  cutoff). 
Because of supersymmetry,
 neither a cosmological term nor a renormalization of Newton's constant
are generated at one-loop, but only
 terms containing at least four derivatives. Thus,  
quantum corrections
to early-time classical behaviour are of relative order $t^{-6}$, 
i.e just like our corrections $(\delta\phi'/\phi')^2$. Note also that
that generating a cosmological constant by quantum corrections would upset
completely the whole PBB scenario.

\end{document}